\documentclass[aps,preprint]{revtex4}%
\usepackage{amsfonts}
\usepackage{amsmath}
\usepackage{amssymb}
\usepackage{graphicx}%
\providecommand{\U}[1]{\protect\rule{.1in}{.1in}}
\begin{document}
\preprint{arXiv:1003.5732}
\title[Short title for running header]{Spectrum of Hydrogen Atom in Space-Time Non-Commutativity}
\author{M. Moumni}
\affiliation{Department of Matter Sciences, University of Biskra, Algeria}
\affiliation{m.moumni@univ-biskra.dz}
\author{A. BenSlama}
\affiliation{Department of Physics, University of Constantine, Algeria}
\affiliation{a.benslama@yahoo.fr}
\author{S. Zaim}
\affiliation{Department of Matter Sciences, University of Batna, Algeria}
\affiliation{zaim69slimane@yahoo.fr}

\begin{abstract}
We study space-time noncommutativity applied to the hydrogen atom and its
phenomenological effects. We find that it modifies the potential part of the
Hamiltonian in such a way we get the Kratzer potential instead of the Coulomb
one and this is similar to add a dipole potential or to consider the extended
charged nature of the proton in the nucleus. By calculating the energies from
the Schr\"{o}dinger equation analytically and computing the fine structure
corrections using perturbation theory, we study the modifications of the
hydrogen spectrum. We find that it removes the degeneracy with respect to both
the orbital quantum number $l$ and the total angular momentum quantum
number\ $j$; it acts here like a Lamb shift. Comparing the results with the
experimental values from spectroscopy, we get a new bound for the space-time
non-commutative parameter. We do the same perturbative calculation for the
relativistic case and compute the corrections of the Dirac energies; we find
that in this case too, the corrections are similar to a Lamb shift and they
remove the degeneracy with respect to $j$ ; we get an other bound for the
parameter of non-commutativity.

\end{abstract}
\volumeyear{year}
\volumenumber{number}
\issuenumber{number}
\eid{identifier}
\date[Date text]{June13,2011}
\received[Received text]{}

\revised[Revised text]{}

\accepted[Accepted text]{}

\published[Published text]{}

\startpage{1}
\endpage{2}
\maketitle
\tableofcontents

\section{Introduction :}

The idea of taking non-commutative space-time coordinates is not new as it
dates from the thirties. It had as objective to regulate the divergences of
quantum field theory by introducing an effective cut-off coming from a
non-commutative structure of space-time at small length scales. Then, it was
abandoned because of problems caused by the violation of unitarity and
causality, but the mathematical development of the theory continued and
especially after the work of Connes in the eighties [1].

In 1999, during their work on string theory, Seiberg and Witten showed that
the dynamics of the endpoints of an open string on a D-brane in the presence
of a magnetic back-ground field is described by a theory of Yang-Mills on a
non-commutative space-time [2]; this has renewed interest in the theory.
Recently, there are many works on Lorentz invariant interpretation of the
theory [3-7].

Today, we find non-commutativity in various fields of physics such as solid
state physics, where it was shown that the Hall conductivity is quantized
within this framework [8] and that non-commutativity is the right tool
replacing Bloch's theory whenever the translation invariance, that occurs in
crystals, is broken in aperiodic solids [9]. Another example is fluid
mechanics, where we define the non-commutative fluids by studying the quantum
Hall effect [10] or bosonization of collective fermion states [11]. One can
also mention the connection with quantum statistical physics [12], the
equivalence between non-commutative quantum mechanics and the Landau problem
in the lowest Landau level [13], the interpretation of Ising-type models as a
kind of field theory in the framework of non-commutative geometry [14] or the
relation with Berry curvature in momentum space [15]. One can even find a
manifestation of the non-commutativity in the physiology of the brain, where
non-commutative computation in the vestibulo-ocular reflex was demonstrated in
a way that is unattainable by any commutative system [16]. The article of
Douglas and Nektasov [17] is an excellent reference for the different
applications of noncommutative field theory.

The theory is a distortion of space-time where the coordinates $x^{%
%TCIMACRO{\U{3bc} }%
%BeginExpansion
\mu
%EndExpansion
}$ become Hermitian operators and thus do not commute:%
\begin{equation}
\left[  x_{nc}^{\mu},x_{nc}^{\nu}\right]  =i\theta^{\mu\nu}=iC^{\mu\nu
}/\left(  \Lambda_{nc}\right)  ^{2};\mu,\nu=0,1,2,3
\end{equation}

The $nc$\ indices denote noncommutative coordinates. $\theta^{%
%TCIMACRO{\U{b5}}%
%BeginExpansion
\mu
%EndExpansion
\nu}$\ is the parameter of the deformation, $C^{%
%TCIMACRO{\U{b5}}%
%BeginExpansion
\mu
%EndExpansion
\nu}$\ are dimensionless parameters and $\Lambda_{nc}$ is the energy scale
where the non-commutative effects of the space-time will be relevant. The
non-commutative parameter is an anti-symmetric real matrix and ordinary
space-time is obtained by making the limit $\theta^{%
%TCIMACRO{\U{b5}}%
%BeginExpansion
\mu
%EndExpansion
\nu}\rightarrow0$. For a review, one can see reference [18].

In the literature, there are a lot of phenomenological studies giving bounds
on the non-commutative parameter. For example, the OPAL collaboration founds
$\Lambda_{nc}\geq140~GeV$ [19], various non-commutative QED processes give the
range $\Lambda_{nc}\geq500~GeV-1.7~TeV$ [20], high precision atomic experiment
on the Lamb shift in the hydrogen atom gives the limit $\Lambda_{nc}^{ss}%
\geq6~GeV$ [21] (This limit corrects the error made in calculating the bound
in [22]); all these bounds deal with space-space non-commutativity. For the
space-time case, the bound $\theta\lesssim9.51\times10^{-18}\ m.s$ was found
from quantum gravity considerations [23], the bound $\theta_{st}%
\lesssim(0.6\ GeV)^{-2}$ was determined in [21] from theoretical limit of the
Lamb shift in H-atom. Some specific models gives the bound $\Lambda_{nc}%
\geq10~TeV$ from CMB data [24] or the bound $\Lambda_{nc}\gtrsim10^{16}%
~GeV$\ from particle phenomenology [25] , but they are not direct constraints
on the parameter because they use the loss of Lorentz invariance in the
theory. A well documented review on non-commutative parameter bounds can be
found in [26].

We are interested in the phenomenological consequences of space-time
non-commutativity. We focus on the hydrogen atom because it is a simple and a
well studied quantum system and so it can be taken as an excellent test for
non-commutative signatures. The case of space-space non-commutativity was
studied by Chaichian et al in [22] and [27]; here we work on the space-time
case. We start by computing the corrections to the Schr\"{o}dinger energies
then we study the contributions to the fine structure. Finally, we compute the
corrections to the Dirac energies and we study the changes in the spectrum.
This allows us to obtain a limit on the non-commutative parameter in each case.

The aim of this work is to study the effects of space-time noncommutativity on
the spectrum of hydrogen atom and to find an upper limit for the
non-commutative parameter by computing the corrections to the transition
energies and comparing with the experimental results from hydrogen spectroscopy.

\section{Hydrogen Atom in Space-Time Non-Commutativity:}

We work here on the space-time version of the non-commutativity; thus instead
of (1), we use:%
\begin{equation}
\left[  x_{st}^{j},x_{st}^{0}\right]  =i\theta^{j0}%
\end{equation}
the $st$ subscripts are for non-commutative space-time coordinates. The $0$
denotes time and $j$ is used for space coordinates. As a solution to these
relations, we choose the transformations:%
\begin{equation}
x_{st}^{j}=x^{j}-i\theta^{j0}\partial_{0}%
\end{equation}
The usual coordinates of space $x^{j}$ satisfy the usual canonical permutation
relations. For convenience we use the vectorial notation:%
\begin{equation}
\overrightarrow{r}_{st}=\overrightarrow{r}-i\overrightarrow{\theta}%
\partial_{0}%
\end{equation}
where we have used the notation:%
\begin{equation}
\overrightarrow{\theta}\equiv\left(  \theta^{10},\theta^{20},\theta
^{30}\right)  =\left(  \theta^{1},\theta^{2},\theta^{3}\right)
\end{equation}
The relations (3) and (4) can be seen as a Bopp's shift [28].

We are dealing with the stationary quantum equations, and this allows us to
consider the energy as a constant parameter. In our computation, we follow the
work done by Chaichian et al. in space-space non-commutativity whatsoever in
the non relativistic case ([22] and [29]) or in the relativistic case [27]; we
use the standard Schr\"{o}dinger and Dirac equations. This is possible because
our choice in the transformations (5) leaves the coordinate $x^{0}$ and all
the momentums $p^{\mu}$ unchanged. One can cite other studies that have used
the same procedure. In the case of non-commutativity being only between time
and space coordinates as is the case in our work, it was shown in [30] that
one has to use the new non-commutative coordinates and momentums instead of
the usual ones in the Schr\"{o}dinger equation. If there is both space-space
and space-time cases of non-commutativity, it has been shown in [23] and [31]
by studying the neutron in the gravitational field, that only the potential
part of the standard equation varies. In a more simple way, we say that the
kinetic energy does not change since it depends on the momentum that remains
unchanged, thus we take the Coulomb potential and construct its
non-commutative image. To achieve this, we have to write the expression of
$r_{st}^{-1}$ :
\begin{equation}
\frac{1}{r_{st}}=\frac{1}{\left\Vert \overrightarrow{r}-i\overrightarrow
{\theta}\partial_{0}\right\Vert }%
\end{equation}
We make the development in series of the expression and because of the
smallness of the non-commutative parameter, as one can see from the bounds
given in the introduction, we restrict ourselves to the $1st$ order in
$\theta$ and neglect the higher order terms:
\begin{equation}
r_{st}^{-1}=\left[  \left(  \overrightarrow{r}-i\overrightarrow{\theta
}\partial_{0}\right)  ^{2}\right]  ^{-1/2}=\left(  1+\frac{i\partial
_{0}\overrightarrow{r}\cdot\overrightarrow{\theta}}{r^{3}}+O(\theta
^{2})\right)
\end{equation}
and thus, one can write the non-commutative Coulomb potential in the
space-time case (up to the $1st$ order $\theta$)\ as follows:
\begin{subequations}
\begin{align}
V_{nc}(r)  &  =-\frac{e^{2}}{r}\left(  1+i\partial_{0}\frac{\overrightarrow
{\theta}\cdot\overrightarrow{r}}{r^{3}}\right) \\
&  =-\frac{e^{2}}{r}-\frac{e^{2}E}{\hbar}\frac{\overrightarrow{\theta}%
\cdot\overrightarrow{r}}{r^{3}}%
\end{align}
where we have used the fact that $i\partial_{0}\psi=H\psi=\left(
E/\hbar\right)  \psi$.

An adequate choice of the parameter is $\overrightarrow{\theta}=\theta
^{r0}\overrightarrow{r}/r$; It is equivalent to write:
\end{subequations}
\begin{equation}
\theta^{\mu\nu}=\left(
\begin{array}
[c]{cccc}%
0 & -\theta & 0 & 0\\
\theta & 0 & 0 & 0\\
0 & 0 & 0 & 0\\
0 & 0 & 0 & 0
\end{array}
\right)
\end{equation}
Here $\mu,\nu=0,1,2,3$ and $\left(  1,2,3\right)  $ means the spherical
coordinates $\left(  r,\vartheta,\varphi\right)  $. This writing is similar to
that in [32] for the case of non-commutative space-time and in [33] for the
space-space case (Another possible choice is $\overrightarrow{\theta}%
=\theta^{30}\overrightarrow{k}$ and we have examined it in [34] in the
non-relativistic case, but it has not been studied in the relativistic case
until now). The choice made in this paper allows us to write the
non-commutative Coulomb potential as (we note $\theta^{r0}=\theta_{st}$):%
\begin{equation}
V_{nc}(r)=-\frac{e^{2}}{r}-\frac{e^{2}E\theta_{st}}{\hbar}\frac{1}{r^{2}%
}+O(\theta_{st}^{2})
\end{equation}
The expression is similar to the Kratzer potential [35]:%
\begin{equation}
V(r)=-e^{2}/r-Ce^{2}/r^{2}%
\end{equation}
(where $C=E\theta_{st}/\hbar$). This kind of potential is introduced to study
the spectrum of the external electron in alkali metals where it is subject to
the effect of the nucleus (Coulomb term) and to the influence of inner
electrons represented by the additional term and which may be interpreted as
the potential of a central dipole; a dipole with its axis directed towards the
electron and thus retains a spherical symmetry[36].

In a same way, the non-commutative Coulomb potential (10) can be interpreted
as the potential energy of a negative charge under the influence of the
superposition of a field produced by a point charge and another field coming
from an electric central dipole; both are placed at the origin.

In other words; the non-commutative Coulomb potential is equivalent to an
electron in a field of a charge distribution whose characteristics are:

- it is not neutral (a positive net charge here) so it gives the usual Coulomb contribution,

- it is not spherically symmetric and this adds the dipole contribution.

Such a distribution exists in the hydrogen atom in the proton; it is an
extended positively charged system composed of three quarks. So applying
space-time non-commutativity to the electron in the hydrogen atom is
equivalent to consider the extended charged nature of the nucleus or the proton.

The fact that the proton has a structure and is a composite particle implies
that non-commutativity cannot be applied to it as for elementary particles
like electron, and it behaves essentially as a commutative particle in the
non-commutative hydrogen atom as mentioned in [28]. This is why we apply
non-commutativity only to the electron in this work.

Now we compute the corrections induced by this additional term in both the
non-relativistic and the relativistic cases.

\subsection{Corrections of the Bohr Energies :}

We work on the standard Schr\"{o}dinger equation and use the non-commutative
Coulomb potential instead of the usual one, because as mentioned above, the
kinetic energy depends on the momentum which remains unchanged and thus does
not change. For this potential, the Schr\"{o}dinger equation is:
\begin{equation}
i\hbar\partial_{0}\psi=-\frac{\hbar^{2}}{2m}\Delta\psi-\frac{e^{2}}{r}\left(
1-i\partial_{0}\frac{\overrightarrow{r}\cdot\overrightarrow{\theta}}{r^{2}%
}\right)  \psi
\end{equation}
We are dealing with the stationary solutions, so we consider the energy as a
constant parameter and write:%
\begin{equation}
H\psi=-\frac{\hbar^{2}}{2m}\Delta\psi-\frac{e^{2}}{r}\left(  1+\frac{E}{\hbar
}\frac{\overrightarrow{r}\cdot\overrightarrow{\theta}}{r^{2}}\right)  \psi
\end{equation}

The spectrum of the Kratzer potential is obtained by defining a new orbital
quantum number:%
\begin{equation}
l_{K}(l_{K}+1)=l(l+1)-2me^{2}C/\hbar^{2}%
\end{equation}
and solving the Schr\"{o}dinger equation with this new number for the Coulomb
potential; by doing this, we obtain the Bohr energies with $n=n_{r}+l_{K}+1$,
and using this transformation, one can easily find the following expression of
the energy (see [36] for example):%
\begin{equation}
E_{n,l}=-\frac{2me^{4}}{\hbar^{2}}\left(  2n-\left(  2l+1\right)
+\sqrt{\left(  2l+1\right)  ^{2}-8mCe^{2}/\hbar^{2}}\right)  ^{-2}%
\end{equation}
Making the replacement $C=E\theta_{st}/\hbar$, we obtain the relation:%
\begin{equation}
E_{n,l}=-\frac{2me^{4}}{\hbar^{2}}\left(  2n-\left(  2l+1\right)
+\sqrt{\left(  2l+1\right)  ^{2}-8mE\theta_{st}e^{2}/\hbar^{3}}\right)  ^{-2}%
\end{equation}
The smallness of the non-commutative parameter allows us to expand the
relation and we restrict ourselves to the $1st$ order in $\theta$. Doing this,
we solve the resulting expression for the energy (we take the energy as a
constant parameter since we are dealing with stationary solutions of the
Schr\"{o}dinger equation) and find:%
\begin{equation}
E_{n,l}=-\frac{me^{4}}{2\hbar^{2}}\frac{1}{n^{2}}\left(  1-\frac{m^{2}e^{6}%
}{\hbar^{5}}\frac{\theta_{st}}{\left(  l+1/2\right)  n^{3}}\right)
\end{equation}
At the outset, we see that the accidental degeneracy with respect to $l$ is
removed. Noticing that the first term is the Bohr's formula, we conclude that
the space-time non-commutative corrections of the hydrogen energy levels in
the framework of the Schr\"{o}dinger equation are:%
\begin{equation}
\Delta E_{n,l}\left(  nc\right)  =\frac{m^{3}e^{10}}{\hbar^{7}}\frac
{\theta_{st}}{\left(  2l+1\right)  n^{5}}%
\end{equation}
One can obtain a limit for $\theta$ by comparing the corrections to transition
energies obtained using (17) with the experimental results from hydrogen
spectroscopy. We take as test levels, $1S$ and $2S$ because we have the best
experimental precision for the transition between them [37]:%
\begin{equation}
f_{1S-2S}=\left(  2446061102474851\pm34\right)  \ Hz
\end{equation}
The non-commutative correction for this transition writes:%
\begin{equation}
\Delta E_{nc}\left(  1S-2S\right)  =\Delta E_{2,0}-\Delta E_{1,0}=0.969\left(
m^{3}e^{10}/\hbar^{7}\right)  \theta_{st}%
\end{equation}
Comparing with the precision of the experimental value in (19), we obtain:%
\begin{equation}
\theta_{st}\lesssim1.05\cdot10^{-19}\ eV^{-2}\approx(3\ GeV)^{-2}%
\end{equation}
This value is better than the limit obtained in both [21] and [23] and it
justifies our expansion of (16) to obtain the energy. If one consider the
limit for the accuracy of the resonance frequency measurements for this
transition found to be $\sim10^{-5}\ Hz$ in [38], then the new bound will be
$\sim(10\ TeV)^{-2}$.

\subsection{Corrections of the Fine Structure Expression :}

The fine structure Hamiltonian of the hydrogen can be written as:%
\begin{equation}
H_{fs}=\frac{p^{2}}{2m}+V-\frac{p^{4}}{8m^{3}c^{2}}+\frac{\hbar}{4m^{2}c^{2}%
}\overrightarrow{\sigma}\cdot\left(  \overrightarrow{\nabla}V\times
\overrightarrow{p}\right)
\end{equation}
Where the first two terms are the Schr\"{o}dinger Hamiltonian, the third one
is the relativistic correction of the kinetic energy and the last term
represents the spin-orbit contribution (we write his general expression from
the non-relativistic limit of the Dirac equation for convenience as one can
find in [39] and [40] for example). $\sigma$ are Pauli matrices.

The value of $\theta$\ obtained in (21) allows us to consider non-commutative
corrections with perturbation theory; to the $1st$ order in $\theta$, the
corrections of the eigenvalues are:%
\begin{equation}
\Delta E_{n,l}\left(  nc\right)  =\left\langle n,l,m_{l}\left\vert
H_{fs}\right\vert n,l,m_{l}\right\rangle
\end{equation}
We follow [40] and [41] and neglect the changes of the eigenvectors; so from
the Schr\"{o}dinger equation, we can write the relativistic correction of the
kinetic energy as:%
\begin{equation}
\frac{p^{4}}{8m^{3}c^{2}}=\frac{1}{2mc^{2}}\left(  \frac{p^{2}}{2m}\right)
^{2}=\frac{1}{2mc^{2}}\left(  E-V\right)  ^{2}%
\end{equation}
We use the relationship $\overrightarrow{\nabla}V=\left(  \overrightarrow
{r}/r\right)  V%
%TCIMACRO{\U{b4}}%
%BeginExpansion
\acute{}%
%EndExpansion
(r)$ that applies for spherical potentials like our non-commutative Coulomb
potential and write:%
\begin{equation}
H_{fs}=\frac{p^{2}}{2m}+V-\frac{1}{2mc^{2}}\left(  E-V\right)  ^{2}+\frac
{1}{4m^{2}c^{2}r}\frac{dV}{dr}\overrightarrow{s}\cdot\overrightarrow{l}%
\end{equation}
where we have used the fact that $\overrightarrow{r}\times\overrightarrow
{p}=\overrightarrow{l}$ and $\hbar\overrightarrow{\sigma}=\overrightarrow{s}$.
Now, we develop (25) using the expression of the potential from (10) and we
find:%
\begin{equation}
H_{fs}=H_{fs}^{\left(  Coulomb\right)  }+\left[  -\frac{e^{2}}{r^{2}}-\frac
{1}{mc^{2}}\left(  E\frac{e^{2}}{r^{2}}+\frac{e^{4}}{r^{3}}\right)
+\frac{\overrightarrow{s}\cdot\overrightarrow{l}}{2m^{2}c^{2}}\left(
\frac{e^{2}}{r^{4}}\right)  \right]  C
\end{equation}
where we have considered the usual fine structure Hamiltonian for the habitual
Coulomb potential:%
\begin{equation}
H_{fs}^{\left(  Coulomb\right)  }=\frac{p^{2}}{2m}-\frac{e^{2}}{r}-\frac
{1}{2mc^{2}}\left(  E^{2}+\frac{e^{4}}{r^{2}}+\frac{2Ee^{2}}{r}\right)
+\frac{e^{2}}{4m^{2}c^{2}}\frac{\overrightarrow{s}\cdot\overrightarrow{l}%
}{r^{3}}%
\end{equation}
and the new non-commutative correction to this Hamiltonian is ($C=E\theta
_{st}/\hbar$):%
\begin{equation}
H_{fs}^{\left(  nc\right)  }=\left[  -\frac{Ee^{2}}{\hbar r^{2}}-\frac
{E}{\hbar mc^{2}}\left(  \frac{Ee^{2}}{r^{2}}+\frac{e^{4}}{r^{3}}\right)
+\frac{Ee^{2}}{2\hbar m^{2}c^{2}}\frac{\overrightarrow{s}\cdot\overrightarrow
{l}}{r^{4}}\right]  \theta_{st}%
\end{equation}
The fine structure Coulomb energies can be found in the literature (For
example [40] or [41]):%
\begin{equation}
E_{n,j}=-\frac{me^{4}}{2\hbar^{2}}\frac{1}{n^{2}}\left[  1+\frac{\alpha^{2}%
}{n}\left(  \frac{1}{j+1/2}-\frac{3}{4n}\right)  \right]
\end{equation}
$\alpha=e^{2}/\hbar c$ is the fine structure constant and $j=l\pm1/2$\ is the
quantum number associated to the total angular momentum $\overrightarrow
{j}=\overrightarrow{l}+\overrightarrow{s}$. The non-commutative correction to
this energies are:%
\begin{equation}
\left\langle H_{fs}^{\left(  nc\right)  }\right\rangle =\left[  -\frac{Ee^{2}%
}{\hbar}\left\langle \tfrac{1}{r^{2}}\right\rangle -\frac{Ee^{2}}{\hbar
mc^{2}}\left(  E\left\langle \tfrac{1}{r^{2}}\right\rangle +e^{2}\left\langle
\tfrac{1}{r^{3}}\right\rangle -\frac{1}{2m}\left\langle \tfrac{\overrightarrow
{s}\cdot\overrightarrow{l}}{r^{4}}\right\rangle \right)  \right]  \theta_{st}%
\end{equation}
To compute these terms, we use the Kramer's recursive relations:
\begin{subequations}
\begin{gather}
\left\langle \frac{1}{r^{2}}\right\rangle =\left(  \frac{1}{a_{0}}\right)
^{2}\frac{1}{n^{3}(l+1/2)}\\
\left\langle \frac{1}{r^{3}}\right\rangle =\left(  \frac{1}{a_{0}}\right)
^{3}\frac{1}{n^{3}(l+1/2)(l+1)l}\\
\left\langle \frac{1}{r^{4}}\right\rangle =\left(  \frac{1}{a_{0}}\right)
^{4}\frac{3n^{2}-l(l+1)}{2n^{5}(l+3/2)(l+1)(l+1/2)l(l-1/2)}%
\end{gather}
where $a_{0}=\hbar^{2}/me^{2}$\ is the $1st$ Bohr radius. Using the expression
of the Bohr energies $E=-me^{4}/2\hbar^{2}n^{2}$\ and writing:
\end{subequations}
\begin{equation}
\left\langle \frac{\overrightarrow{s}\cdot\overrightarrow{l}}{r^{4}%
}\right\rangle =\left\langle \frac{\hbar^{2}}{2r^{4}}\left(  \overrightarrow
{j}^{2}-\overrightarrow{l}^{2}-\overrightarrow{s}^{2}\right)  \right\rangle
=\frac{\hbar^{2}K}{2}\left\langle \frac{1}{r^{4}}\right\rangle \nonumber
\end{equation}
with $K=j\left(  j+1\right)  -l\left(  l+1\right)  -s\left(  s+1\right)
$,\ we find:%
\begin{equation}
\left\langle H_{fs}^{\left(  nc\right)  }\right\rangle =\left\langle
H_{0}^{\left(  nc\right)  }\right\rangle +\left\langle H_{\alpha^{2}}^{\left(
nc\right)  }\right\rangle
\end{equation}
where:%
\begin{subequations}
\begin{gather}
\left\langle H_{0}^{\left(  nc\right)  }\right\rangle =\left(  \frac
{m^{3}e^{10}}{\hbar^{7}}\right)  \frac{\theta_{st}}{2n^{5}(l+1/2)}\\
\left\langle H_{\alpha^{2}}^{\left(  nc\right)  }\right\rangle =\left\langle
H_{0}^{\left(  nc\right)  }\right\rangle \alpha^{2}\left[
\begin{array}
[c]{c}%
\frac{-1}{2n^{2}}\left(  1-\frac{K}{4(l+3/2)(l-1/2)}\right)  +\\
\frac{1}{(l+1)l}\left(  1-\frac{3K}{8(l+3/2)(l-1/2)}\right)
\end{array}
\right]
\end{gather}
The first term is the non-commutative correction found in (17) and the last
terms are the fine structure non-commutative corrections and are proportional
to $\alpha^{2}$ as it should. We have two cases for the value of the orbital
quantum number $l$\ since it can be zero or not, and because $s=1/2$, this
gives three possibilities for $j$.\ We will study the two cases $j=l\pm1/2$
corresponding for $l\neq0$ first then we will treat the case $j=1/2$ when
$l=0$\ separately.

The case $j=l+1/2$ and $l\neq0$:
\end{subequations}
\begin{equation}
\left\langle H_{fs}^{\left(  nc\right)  }\right\rangle _{n,j^{+}}%
=\frac{\left(  \frac{m^{3}e^{10}}{\hbar^{7}}\right)  \theta_{st}}{2n^{5}%
j}\left\{  1+\frac{\alpha^{2}/4}{\left(  j^{2}-1\right)  }\left[
\frac{-8j^{2}+2j+7}{4n^{2}}+\frac{16j^{2}-6j-13}{\left(  4j^{2}-1\right)
}\right]  \right\}
\end{equation}

The case $j=l-1/2$:%
\begin{equation}
\left\langle H_{fs}^{\left(  nc\right)  }\right\rangle _{n,j^{-}}%
=\frac{\left(  \frac{m^{3}e^{10}}{\hbar^{7}}\right)  \theta_{st}}{2n^{5}%
(j+1)}\left\{  1+\frac{\alpha^{2}/4}{j\left(  j+2\right)  }\left[
\frac{-8j^{2}-18j-3}{2n^{2}}+\frac{16j^{2}+38j+9}{(2j+1)(2j+3)}\right]
\right\}
\end{equation}
We see that the expression depends on the way to obtain the number $j$\ from
$l$\ unlike the usual fine structure correction in (28) which is the same for
all the possible values of $j$. It implies that the non-commutativity removes
the degeneracy $j=l+1/2=(l+1)-1/2$ in the hydrogen atom, in states like
$nP_{3/2}$ and $nD_{3/2}$ (the same degeneracy exists in the Dirac energies
for Coulomb potential).

The case $l=0$\ must be treated separately because $l$ appears in the
denominator of two terms in (33) and this gives a divergent result. This
singularity is only apparent because the terms come from the fact that we have
neglected the potential energy of the electron with respect to $mc^{2}$, but
this is not true at the limit $r\rightarrow0$. The same thing appears in the
calculation of the spin-orbit term in usual hydrogen fine structure. It has
been showed in [39] that the spin orbit term is zero in this case because in
the $\left\langle \frac{\overrightarrow{s}\cdot\overrightarrow{l}}{r^{4}%
}\right\rangle $ term, the numerator $\overrightarrow{s}\cdot\overrightarrow
{l}$ vanishes exactly for $l=0$ while the denominator $\left\langle
r^{-4}\right\rangle $ approaches the limit zero. Things are not the same for
the $\left\langle \frac{1}{r^{3}}\right\rangle $ term as there is no vanishing
numerator here and one has to be replace $\left\langle r^{-3}\right\rangle $
by $\left\langle r^{-3}\left(  1-a_{0}\alpha^{2}/2r\right)  \right\rangle $ by
using $\left(  2mc^{2}+e^{2}/r\right)  $\ instead of $\left(  2mc^{2}\right)
$ [39]; this gives us a finite result. We use the same argument to eliminate
the spin-orbit term and the same transformation to compute $\left\langle
r^{-3}\right\rangle $; we get:%
\begin{equation}
\left\langle H_{fs}^{\left(  nc\right)  }\right\rangle _{n,l=0}=\frac
{m^{3}e^{10}}{\hbar^{7}n^{2}}\theta_{st}\left[  \frac{1}{n^{3}}-\frac
{\alpha^{2}}{2}\left(  \frac{1}{n^{5}}-\left\langle r^{-3}\left(
1-\frac{a_{0}\alpha^{2}}{2r}\right)  \right\rangle a_{0}^{3}\right)  \right]
\end{equation}
As an example, we compute the corrections of the states $1s$ and $2s$ :
\begin{align}
\left\langle H_{\alpha^{2}}^{\left(  nc\right)  }\right\rangle _{1s}  &
=-\frac{m^{3}e^{10}}{\hbar^{7}}\theta_{st}\left[  \frac{\alpha^{2}}{2}\left(
1-37.053\right)  \right]  =-\left(  0.96\cdot10^{-3}\right)  \frac{m^{3}%
e^{10}}{\hbar^{7}}\theta_{st}\\
\left\langle H_{\alpha^{2}}^{\left(  nc\right)  }\right\rangle _{2s}  &
=-\frac{m^{3}e^{10}}{\hbar^{7}}\theta_{st}\left[  \frac{\alpha^{2}}{8}\left(
\frac{1}{32}-4.603\right)  \right]  =-\left(  0.03\cdot10^{-3}\right)
\frac{m^{3}e^{10}}{\hbar^{7}}\theta_{st}\nonumber
\end{align}
These terms are added to (18). We see that the additional contribution to the
transition correction (20) is $10^{-3}$ smaller and this do not affect the
limit (21).

We can also compute the correction to the Lamb shift $2P_{1/2}\longrightarrow
2S_{1/2}$ as an example. From (17), (35) and (37), we have:
\begin{subequations}
\begin{gather}
\left\langle H_{fs}^{\left(  nc\right)  }\right\rangle _{2P_{1/2}%
}=0.010\left(  m^{3}e^{10}/\hbar^{7}\right)  \theta_{st}\\
\left\langle H_{fs}^{\left(  nc\right)  }\right\rangle _{2S_{1/2}%
}=0.031\left(  m^{3}e^{10}/\hbar^{7}\right)  \theta_{st}%
\end{gather}
and the Lamb Shift correction follows:%

\end{subequations}
\begin{equation}
\Delta E_{nc}\left(  2P_{1/2}\longrightarrow2S_{1/2}\right)  =0,021\left(
m^{3}e^{10}/\hbar^{7}\right)  \theta_{st}%
\end{equation}
We compare this result to the current theoretical accuracy $0.08\ kHz$ from
[42] and find the bound $\theta_{st}\lesssim(0.4\ GeV)^{-2}$ which is bigger
than the precedent one in (21).

\subsection{Corrections of the Dirac Energies :}

For the relativistic case, we write the Dirac equation:%
\begin{equation}
i\hbar\partial_{0}=H\psi=\left(  \overrightarrow{\alpha}\cdot\overrightarrow
{p}\right)  +m\gamma^{0}+eA_{0}%
\end{equation}
where $\alpha_{i}=\gamma_{0}\gamma_{i}$ and $\gamma_{\mu}$\ are the Dirac matrices.

We use the same argument as in the non-relativistic case and employ the
standard Dirac equation but with the non-commutative Coulomb potential:%
\begin{equation}
A_{0}^{(nc)}=-\frac{e}{r_{st}}=-e\left[  \left(  \overrightarrow
{r}-i\overrightarrow{\theta}\partial_{0}\right)  ^{2}\right]  ^{-1/2}%
=-\frac{e}{r}\left(  1+\frac{i\partial_{0}\overrightarrow{r}\cdot
\overrightarrow{\theta}}{r^{2}}+O(\theta^{2})\right)
\end{equation}
We restrict ourselves to the $1st$ order in $\theta$ and neglect the higher
order terms in the development in series of the expression:%

\begin{equation}
A_{0}^{(nc)}=-\frac{e}{r}\left(  1+i\partial_{0}\frac{\theta_{st}}{r}\right)
+O(\theta^{2})=-\frac{e}{r}-\frac{eE\theta_{st}}{\hbar}\frac{1}{r^{2}%
}+O(\theta^{2})
\end{equation}
The Hamiltonian can now be expressed as:%
\begin{equation}
H=\left(  \overrightarrow{\alpha}\cdot\overrightarrow{p}\right)  +m\gamma
^{0}-e\left(  e/r+e\left(  E/\hbar\right)  \theta_{st}/r^{2}\right)
=H_{0}+H_{nc}%
\end{equation}
$H_{0}$ is the Dirac Hamiltonian in the usual relativistic theory and $H_{nc}
$ is the non-commutative correction to this Hamiltonian:%
\begin{equation}
H_{nc}=-e^{2}\left(  E/\hbar\right)  \theta_{st}r^{-2}%
\end{equation}

The smallness of the parameter $\theta$\ from the different bounds mentioned
in the introduction allows us to consider noncommutative corrections with
perturbation theory; to the $1st$ order in $\theta$, the corrections of the
eigenvalues are:%
\begin{equation}
\Delta E_{nc}=\left\langle H_{nc}\right\rangle =-\left(  Ee^{2}\theta
_{st}/\hbar\right)  \left\langle r^{-2}\right\rangle
\end{equation}

From [43], one has:%
\begin{equation}
\left\langle \frac{1}{r^{2}}\right\rangle =\frac{2\kappa\left(  2\kappa
\varepsilon-1\right)  \left(  1-\varepsilon^{2}\right)  ^{3/2}}{\alpha
\sqrt{\kappa^{2}-\alpha^{2}}\left[  4\left(  \kappa^{2}-\alpha^{2}\right)
-1\right]  }\left(  \frac{mc}{\hbar}\right)  ^{2}\left(  \frac{1}{a_{0}%
}\right)  ^{2}%
\end{equation}
where $a_{0}=\hbar^{2}/me^{2}$\ is the $1st$ Bohr radius and $\varepsilon
=E/mc^{2}$; $E$\ is the Dirac energy:%
\begin{equation}
E_{n,j}=mc^{2}\left\{  1+\alpha^{2}\left[  \left(  n-j-1/2\right)
+\sqrt{\left(  j+1/2\right)  ^{2}-\alpha^{2}}\right]  ^{-2}\right\}  ^{-1/2}%
\end{equation}
$\alpha=e^{2}/\hbar c$ is the fine structure constant and $j=l\pm1/2$\ is the
quantum number associated to the total angular momentum $\overrightarrow
{j}=\overrightarrow{l}+\overrightarrow{s}$. The number $\kappa$\ is giving by
the two relations $\kappa=-\left(  j+1/2\right)  $ if $j=\left(  l+1/2\right)
$ and $\kappa=\left(  j+1/2\right)  $ if $j=\left(  l-1/2\right)  $. We see
that through $\kappa$, the energy depends not only on the value of $j$ but
also on the manner to get this value (or on $l$), unlike the usual Dirac
energies in (48) which is the same for all the possible ways to obtain $j$.
This implies that the non-commutativity removes the degeneracy
$j=l+1/2=(l+1)-1/2$ in hydrogen atom ($nP_{3/2}$ and $nD_{3/2}$ for example)
and acts like the Lamb shift.

We recall that the energy level without considering the rest mass energy
($mc^{2}$) is written as a function of the total energy by the relation
$E_{n,j}=E-mc^{2}$ and so the corrections to these energies are: $\Delta
E_{n,j}^{\left(  nc\right)  }=\Delta E^{\left(  nc\right)  }$. From now on, we
note these corrections $E_{n,j}^{\left(  nc\right)  }$ or $E^{(nc)}\left(
nL_{j}\right)  $ where $L$ is the spectroscopic letter corresponding to a
specific value of the angular quantum number $l$. We take as an example the
levels $n=1,2$ and we compute the corrections to their energies:
\begin{subequations}
\begin{gather}
E^{(nc)}\left(  1S_{1/2}\right)  =1.065084\cdot10^{-4}\left(  m^{3}e^{2}%
c^{4}/\hbar^{3}\right)  \theta_{st}\\
E^{(nc)}\left(  2S_{1/2}\right)  =1.331426\cdot10^{-5}\left(  m^{3}e^{2}%
c^{4}/\hbar^{3}\right)  \theta_{st}\\
E^{(nc)}\left(  2P_{1/2}\right)  =0.443805\cdot10^{-5}\left(  m^{3}e^{2}%
c^{4}/\hbar^{3}\right)  \theta_{st}\\
E^{(nc)}\left(  2P_{3/2}\right)  =0.443765\cdot10^{-5}\left(  m^{3}e^{2}%
c^{4}/\hbar^{3}\right)  \theta_{st}%
\end{gather}

We can get a limit for $\theta$ by comparing these shifts to experimental
results from hydrogen spectroscopy. We take as test levels, $1S-2S$ transition
(19) and looking at (49), the non-commutative correction for this transition
is:
\end{subequations}
\begin{align}
\Delta E^{\left(  nc\right)  }\left(  1S-2S\right)   &  =E^{(nc)}\left(
1S_{1/2}\right)  -E^{(nc)}\left(  2S_{1/2}\right) \nonumber\\
&  =0.931941\cdot10^{-4}\left(  m^{3}e^{2}c^{4}/\hbar^{3}\right)  \theta_{st}%
\end{align}
Comparing with the precision of the experimental value in (15), we obtain:%
\begin{equation}
\theta_{st}\lesssim3.099\cdot10^{-24}\ eV^{-2}\approx(0.57\ TeV)^{-2}%
\end{equation}
It is a significant improvement of the previous bounds obtained in [22] [23]
and [34] and it justifies the use of perturbation method to obtain the energy
. If one consider the limit for the accuracy of the resonance frequency
measurements for this transition found to be $\sim10^{-5}\ Hz$ in [38], then
the new bound will be $\approx$ $(1.5\ PeV)^{-2}$.

To make the differences between the corrections of the levels more visible, we
can write their expressions in a more elegant and appropriate way by using the
development in series with respect to $\alpha$ (up to the $2nd$ order in
$\alpha^{2}$ to do the comparison with the fine structure corrections). For
the Dirac energies, we have:%
\begin{equation}
E=mc^{2}\left\{  1-\frac{\alpha^{2}}{2n^{2}}\left[  1+\left(  \frac{2}{\left(
2j+1\right)  n}-\frac{3}{4n^{2}}\right)  \alpha^{2}\right]  +O(\alpha
^{6})\right\}
\end{equation}
We use this formula and the general expressions from (46) and (47) to compute
the non-commutative corrections and find:
\begin{subequations}
\begin{gather}
E_{n,j=l+\frac{1}{2}}^{\left(  nc\right)  }=\tfrac{m^{3}e^{2}c^{4}\alpha^{2}%
}{jn^{3}\hbar^{3}}\left[  1+\left(  \tfrac{6j^{2}+6j+1}{j\left(  j+1\right)
\left(  2j+1\right)  ^{2}}+\tfrac{3}{\left(  2j+1\right)  n}-\tfrac
{10j+9}{4\left(  j+1\right)  n^{2}}\right)  \alpha^{2}\right]  \theta_{st}\\
E_{n,j=l-\frac{1}{2}}^{\left(  nc\right)  }=\tfrac{m^{3}e^{2}c^{4}\alpha^{2}%
}{(j+1)n^{3}\hbar^{3}}\left[  1+\left(  \tfrac{6j^{2}+6j+1}{j\left(
j+1\right)  \left(  2j+1\right)  ^{2}}+\tfrac{3}{\left(  2j+1\right)
n}-\tfrac{10j+1}{4jn^{2}}\right)  \alpha^{2}\right]  \theta_{st}%
\end{gather}
We see that the non-commutativity acts like a Lamb shift and remove the
degeneracy $j=l+1/2=(l+1)-1/2$ in the hydrogen as we have mentioned before. We
note that the relativistic corrections (53a,53b) are not the same as those
found for the fine structure (35,36,37), whereas they coincide for the quantum
theory of hydrogen atom in the usual case. This coincidence is accidental (as
it is the case of Gauss theorem) and is due to the Coulomb potential which is
a special case. So the additional term in $r^{-2}$ breaks the equivalence and
this induces the difference found.

The non-commutative correction to the Lamb shift follows from the previous
expressions:
\end{subequations}
\begin{gather}
\Delta E_{n,j}^{(nc)}(Lamb\ shift)=E_{n,j=l+1/2}^{\left(  nc\right)
}-E_{n,j=(l+1)-1/2}^{\left(  nc\right)  }\nonumber\\
=\frac{m^{3}e^{2}c^{4}\alpha^{2}}{j(j+1)n^{3}\hbar^{3}}\left[  1+\left(
\frac{6j^{2}+6j+1}{j\left(  j+1\right)  \left(  2j+1\right)  ^{2}}+\frac
{3}{\left(  2j+1\right)  n}-\frac{2}{n^{2}}\right)  \alpha^{2}\right]
\theta_{st}%
\end{gather}

To compare with experience, we apply he result to the $n=2$ and $j=1/2$ case
or the $2P_{1/2}\longrightarrow2S_{1/2}$ Lamb shift (the $28cm$ line). From
(53) (or from (49)), we have:
\begin{equation}
\Delta E_{2,1/2}^{\left(  nc\right)  }\left(  2P_{1/2}\longrightarrow
2S_{1/2}\right)  =0.887621\cdot10^{-5}\left(  m^{3}e^{2}c^{4}/\hbar
^{3}\right)  \theta_{st}%
\end{equation}
We compare this result to the current theoretical accuracy $0.08\ kHz$ from
[42] and find the bound $\theta_{st}\lesssim3.254\cdot10^{-23}\ eV^{-2}%
\approx(0.18\ TeV)^{-2}$. It is larger than the previous one in (51) but it is
still better than the ones from [22] [23] and [34].

\section{Conclusion :}

In this work, we look for space-time non-commutative hydrogen atom and induced
phenomenological effects. We found that applying space-time non-commutativity
to the electron in the H-atom modifies the Coulomb potential to give us the
potential of Kratzer. The additional term is proportional to $r^{-2}$ and we
assimilate it to the field of a central dipole. In other words, the action of
space-time non-commutativity is equivalent to consider the extended charged
nature of the proton in the nucleus.

We started by solving the Schr\"{o}dinger equation for this potential; we have
calculated the corrections induced to energy levels by this non-commutative
effect and we find that the non-commutative corrections remove the degeneracy
of the Bohr energies with respect to the orbital quantum number $l$ and the
energies are labelled $E_{n,l}$. by comparing to experimental results from
high precision hydrogen spectroscopy, we get a new bound for the parameter of
non-commutativity (around $(3\ GeV)^{-2}$).

In a second step, we study the contributions to the fine structure and find
that they have no significant effect on the energies. But we found that the
non-commutative Coulomb potential remove the degeneracy $j=l+1/2=(l+1)-1/2$%
\ of the hydrogen fine structure and the energies write $E_{n,j,l}$; the
non-commutativity acts here like a Lamb shift. This is explained by the fact
that Lamb correction can be interpreted as a shift of $r$\ in the Coulomb
potential due to interactions of the bound electron with the fluctuating
vacuum electric field [44], and non-commutativity is also a shift of $r$ as we
can see from the Bopp's shift.

The same thing was done for the relativistic case where by solving the Dirac
equation, we have calculated the corrections induced to energy levels by this
non-commutative effect. By comparison with experimental results, we get a new
bound for the non-commutative parameter (about $(0.57\ TeV)^{-2}$).

The non-commutative corrections to the Dirac theory of hydrogen atom remove
the degeneracy of the Bohr energies with respect to the orbital quantum number
$l$ and also the degeneracy of the Dirac energies\ with respect to the total
angular momentum quantum number $j$ ($j=l+1/2=(l+1)-1/2$), and the energies
are labelled $E_{n,j,l}$. As in the case of the fine structure, the
non-commutativity has an effect similar to that of the Lamb Shift.

Recently, there has been a certain amount of activity around the theme of
cosmological and astrophysical applications of non-commutative geometry models
of particle physics, for example [45-47]. One can study such applications of
non-commutativity via the Lamb shift line and the $2S-1S$ transition and also
via the Lyman-$\alpha$ ray. We draw attention to the fact that $2S-1S$
transition is used in high precision spectroscopy because of the implication
of these measurements on the values of fundamental physical constants like the
fine structure constant $\alpha$ and the Rydberg constant $R$ [37] and in
tests of Lorentz invariance [47]. The possible variation of the fine structure
constant has relation with primordial light nuclei abundance in the early
universe [48], with f(R) theories in Einstein frame and quintessence models
[49] or with the inhomogeneity of the mass distribution in the early universe
and the cosmological constant [50] (One can find a good review in this last
reference). If we take the bound obtained from the Lamb shift theoretical
accuracy $0.08\ kHz$, we find that it corresponds to a shift in the $2S-1S$
transition frequency equal to $\approx0.8\ kHz$. This value is greater than
the experimental accuracy in (19) and thus the space-time non-commutativity
can be tested here (If we consider the bound from the $28cm$ line).

We have to mention that there is another challenge in the study of the
hydrogen atom in the context of space-time non-commutativity, which is to
determine the spectrum for the choice $\overrightarrow{\theta}=\theta
^{30}\overrightarrow{k}$ in the relativistic case. As has been demonstrated in
our article for the non-relativistic case [34], the additional term to the
Coulomb potential is proportional to $\cos\vartheta$ and therefore
non-commutative contributions are no longer diagonals and the perturbation
theory of order one is no longer valid. We need, in this case, to write the
Hamiltonian matrix for corrections of first and second order in $\theta$ and
then compute the eigenvalues for this system; this is in preparation.

\end{document}